\documentclass[twocolumn]{aastex63}
\usepackage{graphicx}

\shorttitle{LcTools II}
\shortauthors{Schmitt, Vanderburg}
\graphicspath{ {./images/} }

\begin{document}


\title{LcTools II: The QuickFind Method for Finding Signals and Associated TTVs in Light Curves from NASA Space Missions}

\author[0000-0002-5034-0949]{Allan R. Schmitt}
\affiliation{Citizen Scientist, 616 W. 53rd. St., Apt. 101, Minneapolis, MN 55419, USA, \href{mailto:aschmitt@comcast.net}{aschmitt@comcast.net}}

\author[0000-0001-7246-5438]{Andrew Vanderburg} \affiliation{Department of Astronomy, The University of Wisconsin-Madison, 475 N. Charter St., Madison, WI 53706, USA}



\begin{abstract}

This paper describes the new QuickFind method in LcTools for finding signals and associated TTVs (Transit Timing Variations) in light curves from NASA space missions.
QuickFind is adept at finding medium to large sized signals (generally those with S/N ratios above 15) extremely fast, significantly reducing overall processing time for a light curve as compared to the BLS detection method \citep{Kovacs}. For example, on the lead author's computer, QuickFind was able to detect both KOI signals for star 10937029 in a 14 quarter Kepler light curve spanning 1,459 days in roughly 2 seconds whereas BLS took about 155 seconds to find both signals making QuickFind in this example about 77 times faster than BLS. This paper focuses on the user interfaces, data processing algorithm, and performance tests for the QuickFind method in LcTools.

\end{abstract}



\section{Introduction} \label{sec:intro} 

LcTools \citep{Schmitt} is a Windows based software system for finding signals and associated TTVs in the light curves for supported missions and High Level Science Products (HLSPs) at MAST\footnote{\url{https://archive.stsci.edu}}. Supported missions include TESS \citep{Ricker}, K2 \citep{Howell}, and Kepler \citep{Koch}. Supported HLSPs include QLP\footnote{\url{https://archive.stsci.edu/hlsp/qlp}} \citep{2020RNAAS...4..204H}, TESS-SPOC\footnote{\url{https://archive.stsci.edu/hlsp/tess-spoc}} \citep{2020RNAAS...4..201C}, and TASOC\footnote{\url{https://archive.stsci.edu/hlsp/tasoc}} \citep{handberg_rasmus_2019_2579846} for the TESS mission and K2SFF\footnote{\url{https://archive.stsci.edu/hlsp/k2sff}} \citep{2014PASP..126..948V} and EVEREST\footnote{\url{https://archive.stsci.edu/hlsp/everest}} \citep{2016AJ....152..100L} for the K2 mission.

Although LcTools may be used by anyone, from novices to professionals, it is primarily intended for advanced citizen scientists, students, and universities.

The LcTools software system consists of four major applications \textbf{--} LcViewer, LcSignalFinder, LcGenerator, and LcReporter.

LcViewer is a multipurpose light curve viewer and signal processor designed to 1) generate, view, edit, and detrend light curves, 2) detect, record, measure, track, locate, query, and display signals, 3) import and display known signals including TOIs, CTOIs, K2OIs, KOIs, and TCEs, 4) detect and record TTVs, 5) phase fold periodic signals, 6) measure time and flux intervals, and 7) query stellar properties.

LcSignalFinder detects and records signals and associated TTVs found in a large set of light curve files which can then be viewed and analyzed with LcViewer. A signal may be recorded for any type of astronomical event, artifact, or anomaly detected in a light curve whether periodic or non-periodic. Both dips and peaks are supported. Examples of transit based signals include planets, eclipsing binaries, trojans, moons, and comets.

LcGenerator builds light curve files in bulk for subsequent use with LcViewer and LcSignalFinder. LcReporter creates an Excel report for the signals recorded by LcViewer and LcSignalFinder.

See the previous LcTools paper \citep{Schmitt} for detail information on each application in the system.

The primary focus of the current paper is on the QuickFind signal detection method and related features which were added to the system after the original LcTools paper was published. Material that was provided in the original paper will not be repeated here. Readers are encouraged to reference the original paper for background information and context.

This paper is organized as follows. Section \ref{sec:finder} describes the changes made to LcSignalFinder and provides a detailed description of the QuickFind data processing algorithm used in the application. Section \ref{sec:viewer} covers the changes made to LcViewer. Section \ref{sec:reporter} describes a new report in LcReporter for listing the signals and TTVs detected by LcSignalFinder. Section \ref{sec:hlsps} covers the new HLSPs and associated data assets supported in LcTools. Section \ref{sec:performance} describes the QuickFind performance tests that were conducted in LcSignalFinder to assess processing speed, signal detection and precision, and TTV detection and precision. Section \ref{sec:summary} provides a summary and concluding remarks.

The material covered below is based on LcTools V18.3 released on November 25, 2020. The LcTools software system is free and can be obtained from the lead author upon request. Hardware and software requirements for running LcTools can be found in Appendix \ref{sec:rtreqs}.



\section{Changes To \large{L\lowercase{c}S\lowercase{ignal}F\lowercase{inder}} \label{sec:finder}}
\subsection{Setting Up a Job} \label{subsec:findersetup}

Changes to the LcSignalFinder window are shown in Figure \ref{fig:LcSignalFinder}. The following fields were added:

\begin{figure*}[ht]
\includegraphics[scale=0.67]{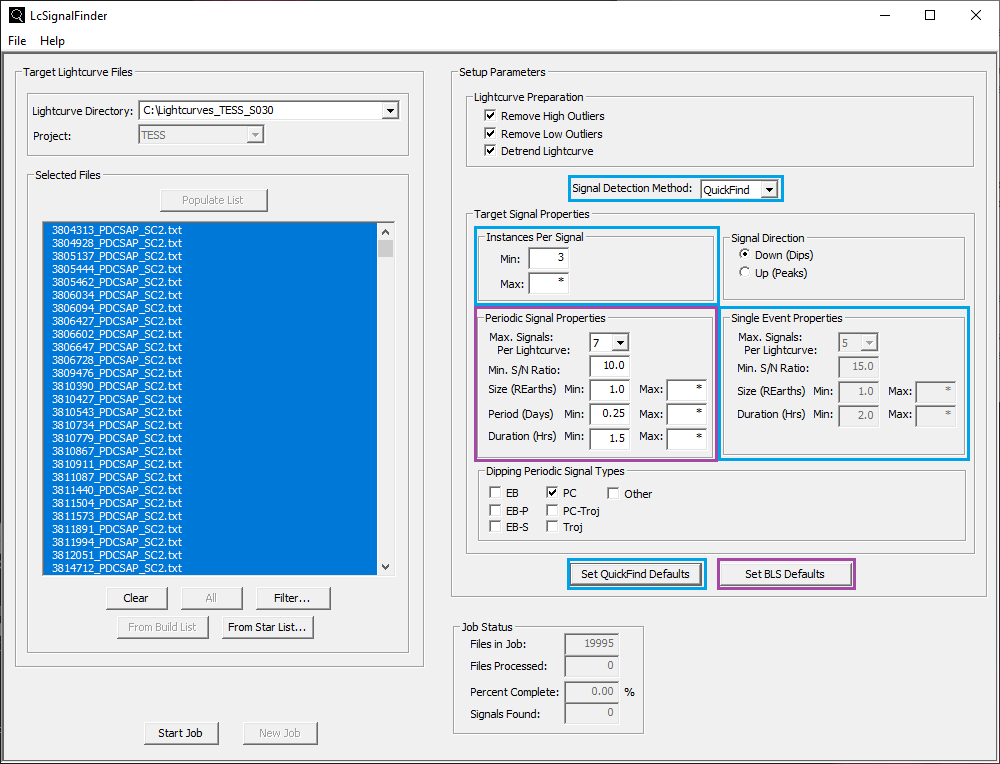}
\centering
\caption{Changes to the LcSignalFinder Window (Blue = Added, Violet = Reorganized or Relabeled).}
\label{fig:LcSignalFinder}
\end{figure*}

\begin{itemize}
\item Signal Detection Method - Sets the target method for finding signals in light curves. Choices are QuickFind and BLS.

\item Instances Per Signal - Sets the minimum and maximum number of instances to find in a signal. A maximum of ``*" indicates unlimited. Setting both fields to 1 will find single events only. Setting both to 2 will find double events only. A minimum of 1 and a maximum of ``*" will find all possible signals in a light curve. A minimum of 3 and a maximum of ``*" will find periodic signals only.

\item Single Event Properties - Sets the target signal properties for single events. Properties include 1) max. number of signals per light curve, 2) min. S/N ratio, 3) min. and max. signal size in REarth units, and 4) min. and max. signal duration in hours. A value of ``*" for max. signal size and max. duration indicates unlimited.

\item Set QuickFind Defaults - Sets default values for the QuickFind method in all fields on the right side of the LcSignalFinder window. The default values are programmed to find periodic signals for planet candidates only.
\end{itemize}

See sections 5.1.1 - 5.1.4 in the previous LcTools paper \citep{Schmitt} for additional information on setting up a job for execution.
\newpage
\subsection{The QuickFind Data Processing Algorithm} \label{subsec:qf}

Once the ``Start Job" button for the QuickFind method is clicked, each selected light curve file on the left side of the LcSignalFinder window will be processed subject to the settings on the right side of the window. Processing of a light curve occurs in three stages -- pre-processing, main processing, and post-processing.

\subsubsection{Stage 1: Pre-Processing} \label{subsec:preproc}

In the pre-processing stage, the application prepares the light curve for analysis. The following steps are performed:

\begin{enumerate}
\item The light curve file is read into memory.
\item Stellar properties for the host star are imported from MAST or NEA\footnote{\url{https://exoplanetarchive.ipac.caltech.edu}} as applicable.
\item Signals for the light curve are imported from the previously selected signal libraries in precedence order and then created in the light curve. Libraries include TOIs, CTOIs, K2OIs, KOIs, TCEs, and user defined signals.
\item TTV offsets for periodic signals are imported from the previously selected TTV libraries in precedence order and then applied to the signals in the light curve to force alignment with the actual signals. Libraries include the Holczer TTV catalog \citep{Holczer_2016} and user defined TTVs.
\item The data points for all project defined and user defined signals in the light curve are removed so that QuickFind will not find and return known signals.
\item Subject to the outlier settings, the worst 1\% of high outliers and the worst 0.05\% of low outliers are removed from the light curve.
\item Subject to the detrend setting, the light curve is detrended using a moving median boxcar size of $\approx$ 21 hours to fit the trend curve.  If a Max. Duration value was specified, the boxcar size will be set to three times the value. This will ensure that the detrending operation will not attenuate (flatten) dips whose duration is less than or equal to the specified value \citep{Hippke}.
\item If the Signal Direction setting is Up, the data points in the light curve are inverted so that peaks become dips enabling QuickFind to detect them.
\end{enumerate}

\subsubsection{Stage 2: Main Processing} \label{subsec:mainproc}
In the main processing stage, the application finds and records all candidate signals and associated TTVs in the light curve. The following actions are performed:

\vspace{4mm} 
\underline{Finding Significant Dips in the Light Curve}

\begin{enumerate}
\item A search is conducted for significant dips in the light curve. On each search iteration, the following steps are performed:
\begin{enumerate}
\item The data point with the greatest depth in the light curve relative to a baseline flux of 1.0 is found. This serves as a rough epoch for a candidate dip.
\item A search region centered at the epoch is determined having a time span of 2.0 days + 3 times the Max. Signal Duration if specified or $\approx$ 21 hours if not.
\item A moving median trend curve is fitted through the data points in the search region using a boxcar size of 3 times the Max. Signal Duration if specified or $\approx$ 21 hours if not.
\item Starting at the epoch, the data points in the search region are scanned backward to find the data point which is closest to the moving median trend curve. This serves as the starting data point for the dip. 
\item Starting at the epoch, the data points in the search region are scanned forward to find the data point which is closest to the moving median trend curve. This serves as the ending data point for the dip.
\item Based on the starting and ending data points for the dip, the dip duration and location are determined.
\item Using a boxcar size of 3 times the dip duration determined in the previous step, a moving median trend curve is fitted through the data points in the search region. Steps d-f are then repeated. This fine tunes the duration and location of the dip.
\item The epoch is reset to the time half-way between the starting and ending data points for the dip.
\item If the S/N ratio of the dip is $\geq$ 6.0, the dip is recorded in a dips table in epoch sequential order. 
\item The data points for the dip are removed from the light curve so that the dip is not detected again.
\item Steps a-j are repeated for the next dip until all significant dips in the light curve are found and recorded.
\end{enumerate}

\underline{Finding Periodic Signals \& TTVs in the Set of Dips}

\item The light curve produced in the pre-processing stage (see section \ref{subsec:preproc}) is restored.
\item For each recorded dip in the dips table, the number of matching dips in the table for the reference dip is determined. Two dips match if their durations and depths are within 20\% of each other.
\item Starting with the reference dip having the greatest number of matching dips in the table, the following steps are performed:

\begin{enumerate}
\item The set of matching dips for the reference dip is obtained.

\item Using the reference dip as the epoch and the time interval between two adjacent matching dips as a trial period, the set of matching dips is examined for a period that overlaps the greatest number of dips and has the highest overlap percent. The periodic signal with the best score is returned.

\item TTVs are calculated for the periodic signal determined in the previous step. First, the signal is created in the light curve based on the reference epoch and period. Then for each each instance of the signal, a search is conducted for the best fitting associated dip in the dips table within a constrained area surrounding the instance. The time difference between the instance and its associated dip is recorded as a TTV.

\item A check is made for outliers in the set of recorded TTVs to help identify inaccurate or erroneous values. An outlier is defined as a TTV value that is at least 1.5 hours greater in magnitude than the TTV value before it and the TTV value after it.

If a TTV outlier is found for an instance, the dips table is searched for another dip that significantly improves or eliminates the outlier. If a better dip is found, the TTV value for the instance is recalculated and re-recorded.

If a better dip is not found, the position of the detected instance is shifted forward or backward in the light curve in an attempt to improve or eliminate the outlier. If a better position is found, the TTV value for the instance is recalculated and re-recorded.

If a better position is not found, the area surrounding the calculated instance is checked for a large data gap in the light curve. If a data gap is found, it is assumed that a suitable dip for the instance could not be found because it resides within the gap. As a consequence, the instance is deemed non-viable and is removed from the fitted results.

\item The periodic signal in the light curve is checked for a smaller period of the form P/N where P is the reference period and N is an integer between 2 and 7. The smallest period is returned.

\item For each instance of the periodic signal, the associated dip in the dips table is flagged as processed so that it is not processed again.

\item Based on the epoch, duration, and period found, a temporary periodic signal is created in the light curve. TTV offsets that were recorded in steps c-d are applied to the signal to align the instances.

\item If the number of instances, S/N ratio, signal size, signal period, and signal duration of the temporary signal matches the settings in the Periodic Signal Properties group box, the periodic signal is recorded in the candidate signals table. The entry is stored in highest-to-lowest S/N ratio order within the periodic group of signals at the top of the table.

\item Steps a-h are then repeated for the next reference dip having the greatest number of matching dips in the table until all reference dips are processed.
\end{enumerate}

\newpage
\underline{Finding Double Events in the Set of Dips}

\item If the Min. Instances Per Signal setting is less than 3, the dips table is scanned for matching unprocessed dips. Two dips match if their durations and depths are within 20\% of each other.
\item For each pair of adjacent unprocessed matching dips in the table, the following steps are performed:
\begin{enumerate}
\item The time interval between matching dips is determined. This serves as the trial period.
\item Based on the epoch of the first dip and the trial period, a test signal is created in the light curve.
\item If the test signal has exactly two instances that overlap the set of unprocessed dips in the table, the test signal is a considered a double event. Each dip in the pair is then flagged as processed in the dips table so that it is not processed again.
\item If the number of instances, S/N ratio, signal size, signal period, and signal duration of the test signal matches the settings in the Periodic Signal Properties group box, the test signal is recorded in the candidate signals table. The entry is stored in highest-to-lowest S/N ratio order within the periodic group of signals at the top of the table.
\item Steps a-d are then repeated for the next pair of adjacent unprocessed matching dips in the table until all pairs are processed.
\end{enumerate}

\underline{Finding Single Events in the Set of Dips}

\item If the Min. Instances Per Signal setting is 1, the dips table is scanned for residual unprocessed dips. If the S/N ratio, signal size, and signal duration of an unprocessed dip matches the settings in the Single Event Properties group box, the dip is recorded in the candidate signals table. The entry is stored in highest-to-lowest S/N ratio order within the non-periodic group of signals at the bottom of the table.
\end{enumerate}

\subsubsection{Stage 3: Post-Processing} \label{subsec:postproc}
In the post-processing stage, the application finishes analyzing and  consolidating the signals recorded in the candidate signals table. The following actions are performed:
\begin{enumerate}
\item For each signal in the table, a preliminary signal type is assigned as follows:

\begin{itemize}
\item If the Signal Direction setting is Up, the signal type is ``Peak". This is true for single events, double events, and periodic signals.
\item If the Signal Direction setting is Down and the signal is a single event, the signal type is ``Dip".
\item If the signal does not have significant TTVs and has a signal duration $>$ 0.2 times the period, the signal type is ``Other".
\item If the estimated object size for the signal is $\geq$ 20 REarths, the signal type is ``EB".
\item If none of the above applies, the default signal type is ``PC".
\end{itemize}

\item For each dipping periodic signal in the table, the signal type is further classified as follows:

\begin{itemize}
\item If there is another dipping periodic signal in the table having a period within 2 hours of the reference period, the two signals are considered an EB pair. The deeper signal is assigned a signal type of ``EB-P" while the shallower signal is assigned a signal type of ``EB-S".
\item If the depths of the even and odd instances of the periodic signal differ by more than 15\%, the signal is considered an EB pair and is split into two signals having a period twice the original. The deeper signal is assigned a signal type of ``EB-P" while the shallower signal is assigned a signal type of ``EB-S".
\item If there is another dipping periodic signal in the table having nearly the same period as the reference signal and whose epoch is offset from the reference epoch by roughly 1/6th of a period (corresponding to the L4 and L5 Lagrange points), the two signals are considered a trojan pair. The deeper signal is assigned a signal type of ``PC-Troj" while the shallower signal is assigned a signal type of ``Troj".
\end{itemize}

\item All duplicate periodic signals are removed from the table.
\item If the Signal Direction setting is Down, all periodic signals whose assigned signal type does not match a selected signal type under the Dipping Periodic Signal Types group box are removed from the table.
\item Based on the Max. Signals Per Lightcurve setting for periodic signals and single events, excess signals are removed from the table starting from the bottom of the applicable signal group. Because signals are retained in highest-to-lowest S/N order within a signal group, the weakest signals are removed first.
\item All surviving signals and associated TTVs are written as text files into the light curve directory for subsequent use by LcViewer and LcReporter.
\end{enumerate}


\section{Changes To \large{L\lowercase{c}V\lowercase{iewer}}} \label{sec:viewer}

\subsection{The ``Setup Work Group" Dialog Box} \label{subsec:workgroup}

A work group is a set of light curve files to be viewed or processed by LcViewer. A work group is defined using the ``Setup Work Group" dialog box. See section 6.2.1 in the previous LcTools paper \citep{Schmitt} for additional information.

Changes to the dialog box are shown in Figure \ref{fig:SetupWorkGroupDlg}. Buttons outlined in blue were added. Buttons outlined in violet were modified. The following changes were made:

\begin{figure}[ht]
\includegraphics[scale=0.75]{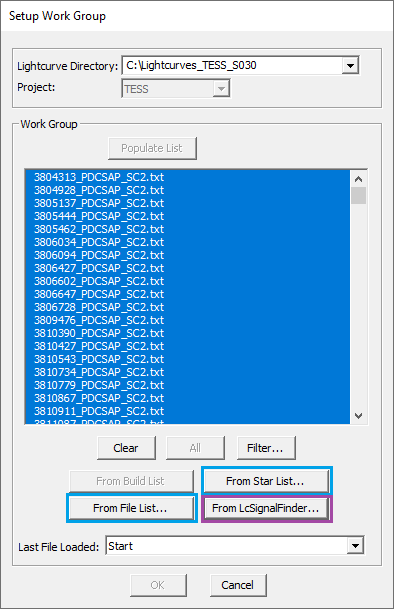}
\centering
\caption{Changes to the ``Setup Work Group" Dialog Box (Blue = Added, Violet = Modified).}
\label{fig:SetupWorkGroupDlg}
\end{figure}

\begin{itemize}
\item Added the ``From Star List" button for selecting  light curve files based on the star IDs found in a specified text file.

\item Added the ``From File List" button for selecting light curve files based on the filenames found in a specified text file.

\item Modified the behavior of the ``From LcSignalFinder" button (originally labeled ``From Signals Found") for selecting light curve files based on the signals detected by LcSignalFinder. Three user options are available:

\begin{enumerate}
\item Select all light curve files having signals from LcSignalFinder.

\item Generate an Excel report of the signals found by LcSignalFinder and then manually select a subset of the light curve files from the report. See Figure \ref{fig:CandSigsRpt} for a sample report.

\item Using the Excel report produced from option number 2, manually select a different subset of light curve files from the report.
\end{enumerate}
\end{itemize}


\subsection{The ``Find Signals" Dialog Box} \label{subsec:findsigsdlg}

The ``Find Signals" dialog box is used to vet the candidate signals recorded by LcSignalFinder for the selected work group. It is also used to record and vet candidate signals detected by LcViewer in the current light curve. See sections 6.13 and 6.14.1 in the previous LcTools paper \citep{Schmitt} for additional information on each feature.

Numerous changes were made to the dialog box as shown in Figure \ref{fig:FindSigsDlg}. Sections outlined in blue were added. Sections outlined in violet were modified. The following changes were made:

\begin{figure*}[ht]
\includegraphics[scale=0.61]{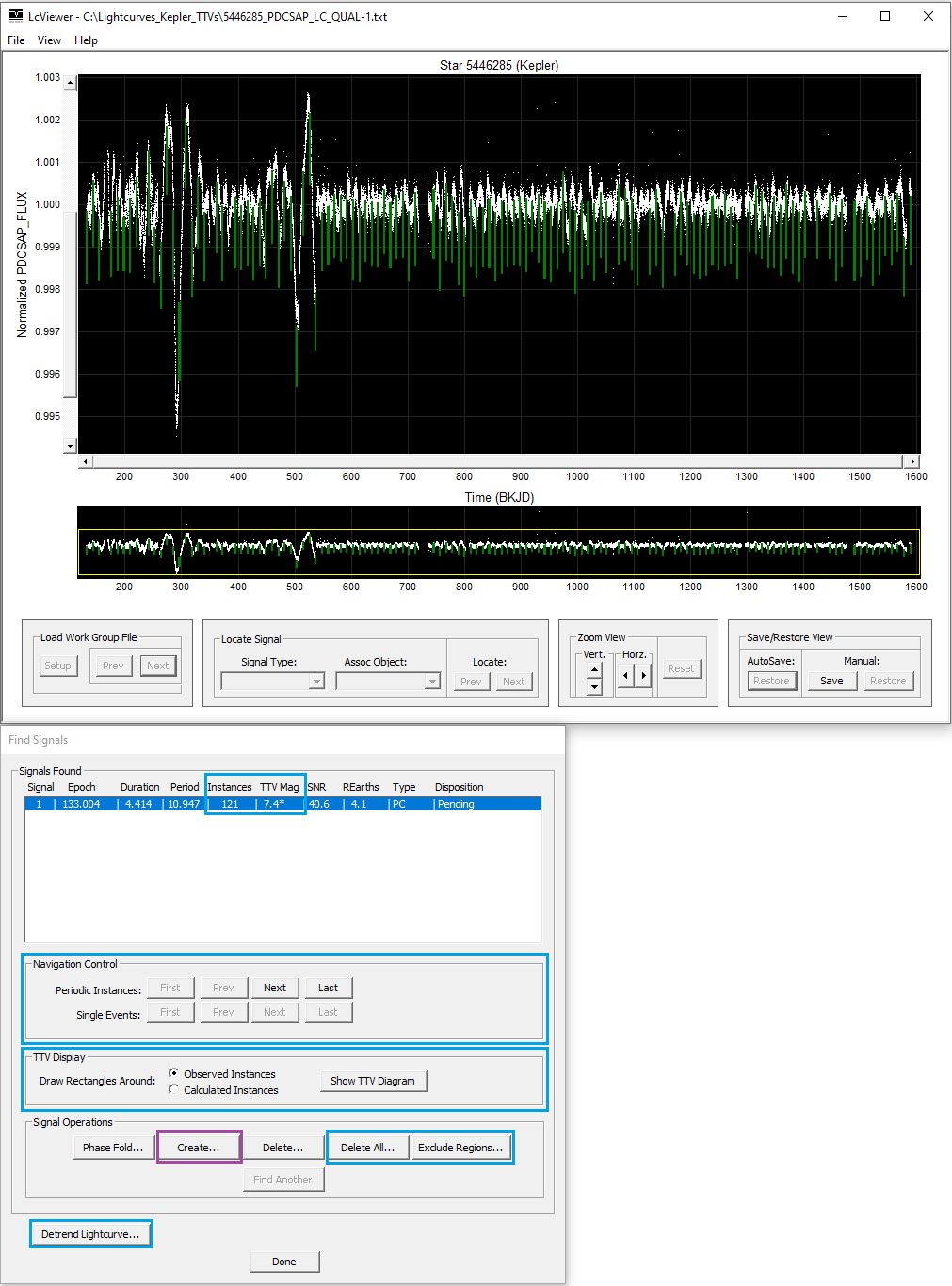}
\caption{Changes to the ``Find Signals" Dialog Box (Blue = Added, Violet = Modified).}
\label{fig:FindSigsDlg}
\end{figure*}

\begin{figure}[ht]
\includegraphics[scale=0.66]{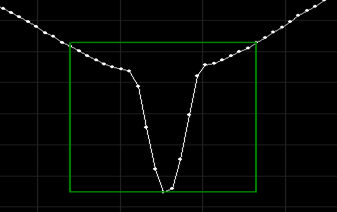}
\centering
\caption{Example of an Underfitted Signal.}
\label{fig:UnderfittedSig}
\end{figure}

\begin{figure}[ht]
\includegraphics[scale=0.66]{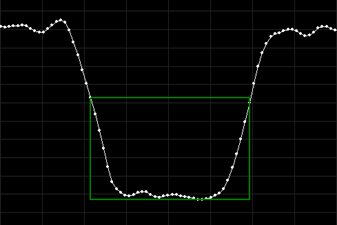}
\centering
\caption{Example of an Overfitted Signal.}
\label{fig:OverfittedSig}
\end{figure}

\begin{itemize}
\item Added the ``Instances" and ``TTV Mag" fields to the signal table at the top of the dialog box. The Instances field shows the number of instances found in a signal where 1 is a single event, 2 is a double event, and $\geq$ 3 is a periodic signal. The TTV Mag field shows the mean TTV magnitude in hours across all instances of a periodic signal. An asterisk (*) is appended to the value if the TTV magnitude is significant.

\item Added the ``Navigation Control" section for manually iterating through the instances of a signal in the light curve. Each instance of the signal is outlined with a green rectangle. The top row of buttons is used navigate through the instances of a periodic signal or double event while the bottom row of buttons is used to navigate though single events.

\item Added the ``TTV Display" section for inspecting a periodic signal with TTVs. 

The radio box on the left side controls how the green rectangles are displayed in the light curve. If the top radio button is selected, the green rectangles are aligned with the observed (i.e., detected) instances -- those with TTV offsets applied. If the bottom radio button is selected, the rectangles are drawn at the calculated instance positions based on the reference epoch and period of the signal. Thus by toggling the radio buttons, it is possible to see how much each instance of the signal was shifted in the light curve due to TTVs.

The ``Show TTV Diagram" button displays a TTV plot for a periodic signal. See section \ref{subsec:ttvdiagram} for details.

\item Changed the behavior of the ``Create" button so that detected TTVs for the selected periodic signal are automatically added to the default TTV library when the signal is saved. See section 6.8.1 in the previous LcTools paper \citep{Schmitt} for additional information on creating a periodic signal.

\item Added the ``Delete All" button for deleting all pending signals from the dialog box and then terminating the Find Signals operation. This is a fast and efficient way to dismiss a large batch of unwanted signals.

\item Added the ``Exclude Regions" button for registering the current signal as a common unwanted signal that should be eliminated from all light curves. Upon clicking this button, each instance of the signal will be recorded in an excluded regions list. See section \ref{subsec:excludedregions} for further information on excluded regions.

\item Added the ``Detrend Lightcurve" button for manually detrending the light curve to obtain a better fit for signals that were underfitted or overfitted as shown in Figures \ref{fig:UnderfittedSig} and \ref{fig:OverfittedSig}.

Upon clicking the ``Detrend Lightcurve" button, the ``Find Signals" dialog box will close and the ``Detrend Lightcurve" dialog box will open for controlling the operation. See section 6.12 in the previous LcTools paper \citep{Schmitt} for how to manually detrend a light curve to obtain an optimal fit. Upon clicking the OK button in the dialog box to complete the operation, LcViewer will automatically search for signals in the detrended light curve using the original settings and then re-open the ``Find Signals" dialog box with the updated results. 

The above procedure is a fast and efficient way to optimize the fit for signals that were underfitted or overfitted in the light curve.

\end{itemize}

\subsection{Moving an Instance of a Periodic Signal}

In QuickFind mode, if the green rectangle for an instance of a periodic signal is misaligned with respect to the actual instance in the light curve, it can be manually aligned using the m+M2 button. Moving the rectangle for an instance of a periodic signal generates a TTV offset record for the signal.

If no actual instance exists at the location, the green rectangle can be deleted from the light curve using the Ctrl+d key.

\subsection{The TTV Diagram (New)} \label{subsec:ttvdiagram}

\begin{figure*}[ht]
\includegraphics[scale=0.66]{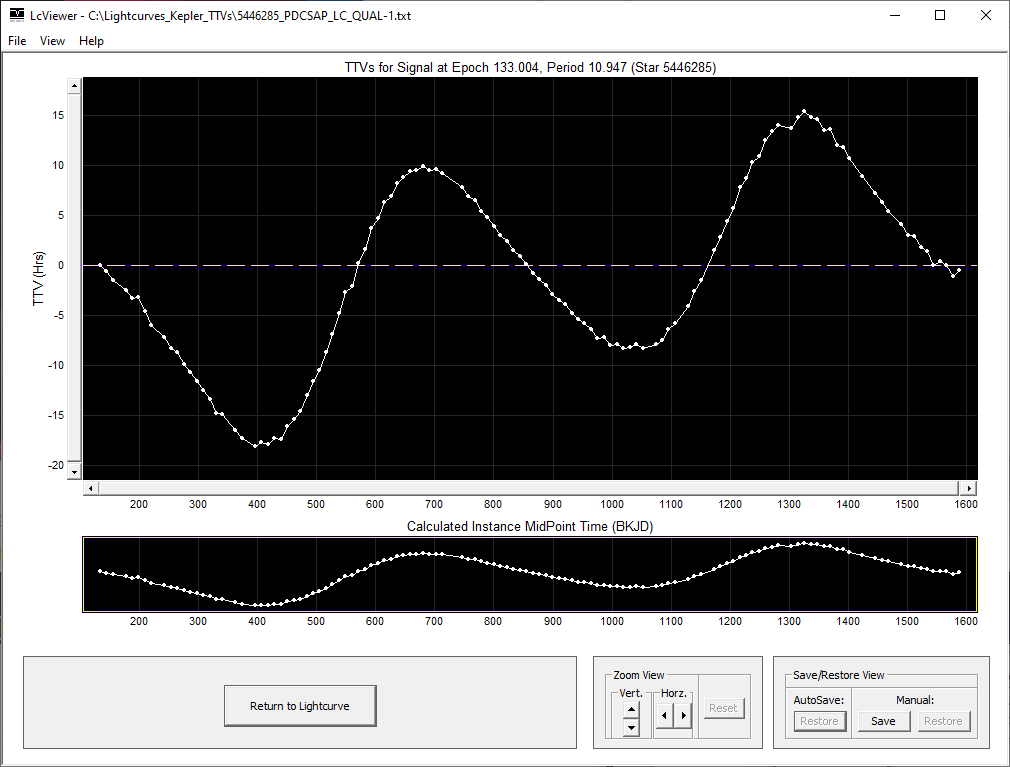}
\caption{TTV Diagram for the Periodic Signal Shown in Figure \ref{fig:FindSigsDlg}.}
\label{fig:TTVDiagram}
\end{figure*}

A TTV diagram (also known as an O-C diagram) shows the TTV value for each instance of a periodic signal in a light curve where a TTV value is defined as the time difference in hours between the (O)bserved instance and the (C)alculated instance in the light curve. Strong TTVs can indicate the presence of another object in the star system whose gravity perturbs the orbit of the host object therefore affecting its timing characteristics.

A TTV diagram can be displayed in one of two ways: 1) by clicking the ``Show TTV Diagram" button in the ``Find Signals" dialog box (see Figure \ref{fig:FindSigsDlg}), or 2) by pressing the Ctrl+t key on an instance of a previously defined periodic signal in the light curve having prerecorded TTVs in a library. A sample TTV diagram for the first method is shown in Figure \ref{fig:TTVDiagram}.

The y-axis shows TTV values in hours while the x-axis shows instance times in project specific days. Each data point in the diagram shows the TTV value at a calculated instance midpoint time. A TTV value can be positive, negative, or zero. It is positive if the observed instance is located after the calculated instance, negative if the observed instance is located before the calculated instance, and zero if no TTV was detected.

User interaction with a TTV diagram is similar to a regular light curve. Navigating within a diagram can be accomplished using the scroll bars, Zoom View controls, or by using the hot keys and mouse buttons listed in the Light Curve Navigation section of Figure \ref{fig:Hotkeys}. The position of the cursor in the diagram can be tracked in real-time using the Tracking Information Box. The interval between two locations in the diagram can also be measured using the Shift+M1 button. Positions and intervals are displayed in time and TTV units rather than in time and flux units as they are for regular light curves. See sections 6.5.1 - 6.5.4 and sections 6.6 - 6.7 in the previous LcTools paper \citep{Schmitt} for additional information on navigation, tracking, and measuring.

By double-clicking the mouse on a data point in the diagram, the diagram will close and the regular light curve will re-open with the view centered on the associated instance in the light curve. This navigation technique is especially useful for quickly correcting misaligned TTVs which manifest as single point outliers in the diagram. 

For example, assume a high data point outlier exists at 200 BKJD in Figure \ref{fig:TTVDiagram}. By double-clicking the mouse on the outlier, the TTV diagram will close and the regular light curve will re-open with the view centered on the instance at 200 BKJD. At that point the user can either move (re-align) the green rectangle with the actual signal using the m+M2 button, delete the rectangle using the Ctrl+d key, or do nothing and accept the TTV results.

\subsection{Excluded Regions (New)} \label{subsec:excludedregions}

An excluded region is a data point region to automatically remove from light curves at load time. A region is defined by a starting and ending time in days. It is typically used to filter out an unwanted bad-spot or signal-like event (periodic or non-periodic) that manifests in the same location across multiple light curves. Using this feature, common unwanted sections of light curves can be eliminated prior to processing them in LcViewer.

\begin{figure}[ht]
\includegraphics[scale=0.67]{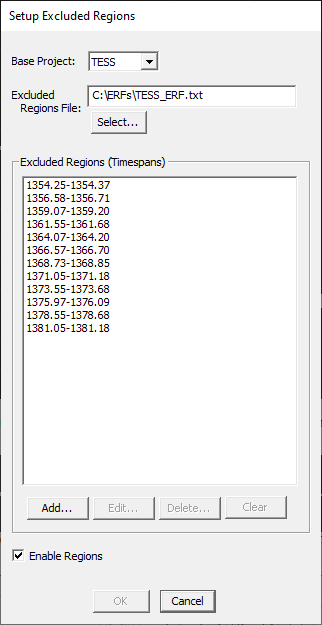}
\centering
\caption{The ``Setup Excluded Regions" Dialog Box.}
\label{fig:SetupExcludedRegionsDlg}
\end{figure}

An excluded region can be recorded in one of two ways: 1) by clicking the ``Exclude Regions" button in the ``Find Signals" dialog box (see Figure \ref{fig:FindSigsDlg}) or 2) by pressing the Shift+e key to open the ``Setup Excluded Regions" dialog box (see Figure \ref{fig:SetupExcludedRegionsDlg}) and then clicking the ``Add" button to add a region to the list. 

\subsection{The ``Find Signals - Setup" Dialog Box} \label{subsec:findsigssetup}

\begin{figure}[ht]
\includegraphics[scale=0.72]{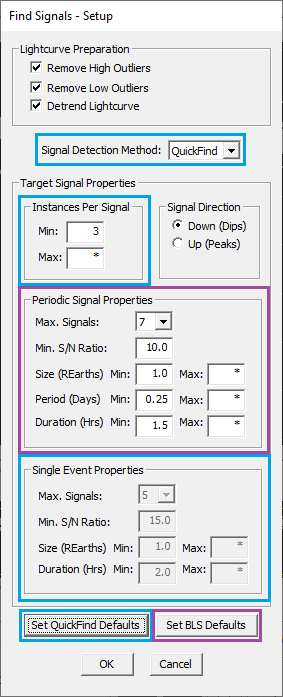}
\centering
\caption{Changes to the ``Find Signals - Setup" Dialog Box (Blue = Added, Violet = Reorganized or Relabeled).}
\label{fig:FindSigsSetupDlg}
\end{figure}

The ``Find Signals - Setup" dialog box is used to set the search parameters for finding signals in the current light curve. Numerous changes were made to the dialog box as shown in Figure \ref{fig:FindSigsSetupDlg}. The changes are virtually identical to the ones for LcSignalFinder as described in section \ref{subsec:findersetup}. As such, they will not be repeated here.

\subsection{The ``Stellar Properties" Dialog Box}

\begin{figure}[ht]
\includegraphics[scale=0.8]{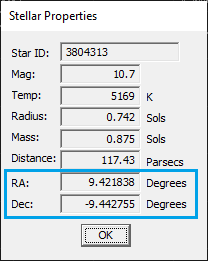}
\centering
\caption{Changes to the ``Stellar Properties" Dialog Box (Blue = Added).}
\label{fig:StellarPropsDlg}
\end{figure}

The RA and Dec fields were added to the ``Stellar Properties" dialog box as shown in Figure \ref{fig:StellarPropsDlg}. The dialog box now shows eight properties: star ID, magnitude, temperature, radius, mass, distance, right ascension, and declination.

\newpage
\subsection{Other Changes}

New menu bar commands and hot keys were added to the LcViewer user interface. See Figures \ref{fig:MenuBar} and \ref{fig:Hotkeys} for details.


\section{Changes To \large{L\lowercase{c}R\lowercase{eporter}}} \label{sec:reporter}

\begin{figure*}[ht]
\includegraphics[scale=0.64]{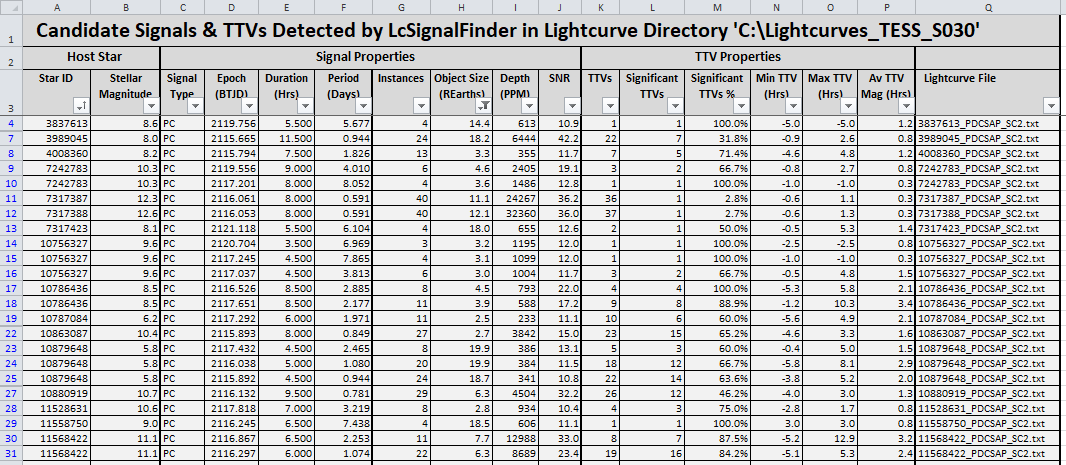}
\centering
\caption{The ``Candidate Signals Report" Produced by LcReporter.}
\label{fig:CandSigsRpt}
\end{figure*}

The ``Candidate Signals Report" was added to LcReporter listing the signals and associated TTVs recorded by LcSignalFinder in a light curve directory. See Figure \ref{fig:CandSigsRpt} for an example.

Each row in the report lists the properties for one signal. Rows are divided into the following sections:
\begin{itemize}
\item Host Star section showing the star ID and stellar magnitude.
\item Signal Properties section showing the signal type, epoch, duration (hrs), period (days), number of instances in the signal, object size (REarths), depth (ppm), and S/N ratio.

\item TTV Properties section showing the number of non-zero TTVs found, number of significant TTVs (those $\geq$ 1 hr), percent of TTVs that are significant, minimum TTV value (hrs), maximum TTV value (hrs), and mean TTV magnitude (hrs).

\item Lightcurve File section showing the name of the light curve file in the directory.
\end{itemize}


\section{New HLSPs and Associated Data Assets} \label{sec:hlsps}
Two new HLSPs for the TESS mission were added to LcTools -- QLP\footnote{\url{https://archive.stsci.edu/hlsp/qlp}} and  TESS-SPOC\footnote{\url{https://archive.stsci.edu/hlsp/tess-spoc}}.

Light curve files for each HLSP can be built via LcGenerator and LcViewer using the star list files provided on the LcTools website\footnote{\url{https://sites.google.com/a/lctools.net/lctools/data-sources}}. Star list files are available for both single sectors and multiple sectors. Roughly 5 million pre-built light curve files for QLP are also available on the website.

See sections 2 and 3 in the previous LcTools paper \citep{Schmitt} for additional information on supported HLSPs and data assets.



\section{Q\lowercase{uick}F\lowercase{ind} Performance Tests} \label{sec:performance}

To assess the performance of the QuickFind method in LcSignalFinder, a number of tests were developed and executed. Tests covered three broad performance categories -- processing speed, signal detection and precision, and TTV detection and precision.

To serve as search targets for the study, KOI signals from the Kepler mission were used. The NASA Exoplanet Archive\footnote{\url{https://exoplanetarchive.ipac.caltech.edu}} was first queried to obtain a list of KOIs and their properties. Properties included the KOI number, host star ID, number of quarters in which the host star was observed, total days observed, signal epoch, duration, period, S/N ratio, and signal disposition. In addition, the Holczer TTV catalog \citep{Holczer_2016} was consulted to obtain the maximum TTV value for each KOI. The above properties were then recorded in a KOI spreadsheet to help guide the study.

All tests were executed using LcTools V18.3 on a Dell XPS 8910 3.4 GHz desktop computer with 16 GB of RAM running under Microsoft Windows 10.

\subsection{Test 1 -- Processing Speed} \label{subsec:processingspeed}

\subsubsection{Objective} \label{subsec:test1objective}

The primary objective of test 1 was to compare the speed of QuickFind with the speed of BLS in processing periodic signals and light curves over various timespans.

Note that this test only measured raw processing speed, completely ignoring quality considerations such as precision. The test was designed to collect all periodic signals in a light curve regardless of viability or signal type.

\subsubsection{Procedure} \label{subsec:test1procedure}

The KOI spreadsheet was first queried for host stars that were observed over a full 17 Kepler quarters (1,459 days). From this list, 17 groups of 25 stars per group (425 stars total) were selected where each group N was assigned to cover quarters 1 thru N. For example, group 7 was assigned to cover quarters 1-7, a timespan of 588 days. The stars in each group were different.

For each group of host stars, LcGenerator was then executed to build Kepler long cadence PDCSAP light curves for the target quarters. To maximize the number of data points in each light curve, the quality filter was turned off.

\begin{figure}[ht]
\includegraphics[scale=0.58]{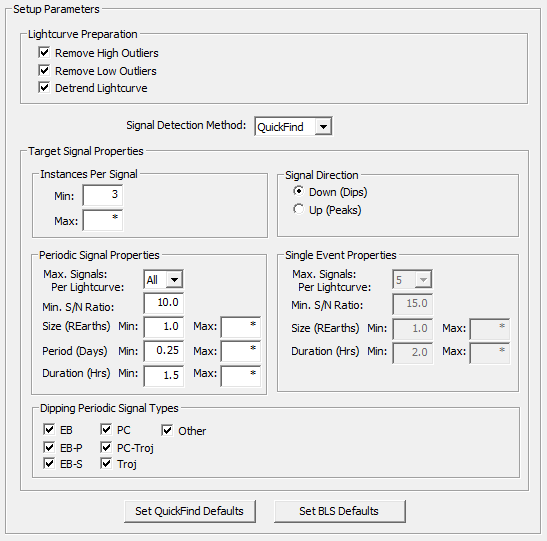}
\centering
\caption{LcSignalFinder Settings Used for QuickFind in Test 1. Except for the signal detection method, BLS used identical settings.}
\label{fig:LcSignalFinderTest1Settings}
\end{figure}

Next, for each group of light curves built with LcGenerator, LcSignalFinder was executed twice -- once for QuickFind and once for BLS. Except for the signal detection method, the settings used for each execution were identical as shown in Figure \ref{fig:LcSignalFinderTest1Settings}.

Prior to starting the job, the signal libraries for KOIs and TCEs were deselected thereby enabling LcSignalFinder to detect those signals.

Upon completion, speed metrics from LcSignalFinder were recorded in an Excel spreadsheet and then graphed. 

\subsubsection{Results} \label{subsec:test1results}

\begin{figure}[ht]
\includegraphics[scale=0.55]{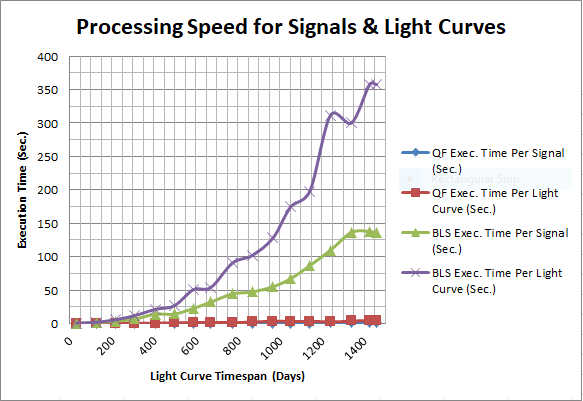}
\centering
\caption{QuickFind (QF) and BLS Processing Speed for Signals and Light Curves Across Various Timespans.}
\label{fig:ProcessingSpeed}
\end{figure}

Figure \ref{fig:ProcessingSpeed} shows the average execution time per signal and the average execution time per light curve for QuickFind and BLS at various timespans. As can be seen, QuickFind was much faster than BLS at all timespans. In general with both methods, as the timespan increased the execution time increased but at a much greater rate with BLS particularly above 300 days.

For QuickFind, the average execution time per signal ranged from 0.14 seconds at 33 days to 1.11 seconds at 1,340 days. The average execution time per light curve ranged from 0.16 seconds at 33 days to 4.72 seconds at 1,459 days.

For BLS, the average execution time per signal ranged from 0.48 seconds at 33 days to 138 seconds at 1,426 days. The average execution time per light curve ranged from 0.44 seconds at 33 days to 358 seconds at 1,426 days.

\begin{figure}[ht]
\includegraphics[scale=0.55]{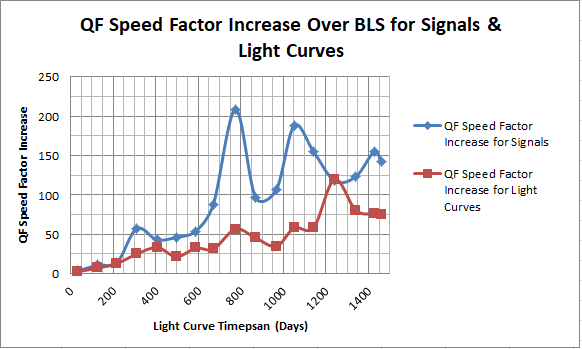}
\centering
\caption{QuickFind (QF) Speed Factor Increase Over BLS for Signals and Light Curves.}
\label{fig:SpeedFactorIncrease}
\end{figure}

To better illustrate the speed difference, Figure \ref{fig:SpeedFactorIncrease} shows the speed factor increase of QuickFind over BLS for signals and light curves at various timespans. For signals, QuickFind was 3.47 times faster than BLS at 33 days, 97 times faster at 869 days, and 142 times faster at 1,459 days. For light curves, QuickFind was 2.75 times faster than BLS at 33 days, 46 times faster at 869 days, and 76 times faster at 1,459 days.

\subsection{Test 2 -- Detection of Periodic Signals} \label{subsec:periodicsignaldetection}

\subsubsection{Objectives} \label{subsec:test2objectives}

The primary objectives of test 2 were to 1) determine the detection rate of KOI signals in a set of light curves over various S/N ratios, and 2) determine the deviation in epoch, duration, and period between the reference KOI signals and the detected KOI signals. 

\subsubsection{Procedure} \label{subsec:test2procedure}

Target KOIs for the test were first selected from the KOI spreadsheet using the data filters TTV value $\leq$ 0.5 hours, S/N ratio $\geq$ 10, signal duration $\geq$ 1.5 hours, and signal disposition = CONFIRMED or CANDIDATE. This produced a initial list of 179 KOIs. Five KOIs were removed from the list due to various issues leaving 174 KOIs for the test.

Next, LcGenerator was executed to build Kepler long cadence PDCSAP light curves for the host stars using all available time series data. To maximize the number of data points in each light curve, the quality filter was turned off.

\begin{figure}[ht]
\includegraphics[scale=0.59]{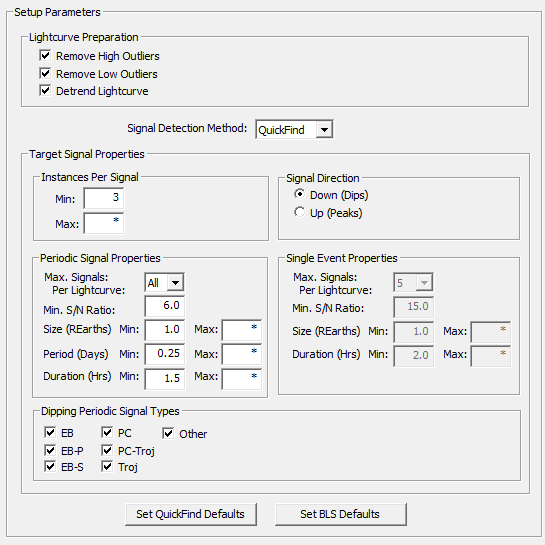}
\centering
\caption{LcSignalFinder Settings Used for Test 2.}
\label{fig:LcSignalFinderTest2Settings}
\end{figure}

LcSignalFinder was then executed on the light curves produced by LcGenerator using the settings shown in Figure \ref{fig:LcSignalFinderTest2Settings}.

Prior to starting the job, the signal library for KOIs was deselected thereby enabling LcSignalFinder to detect KOI signals.

Upon completion, LcReporter was executed to generate an Excel report listing the signals detected by LcSignalFinder. A custom script was then run which combined the reference KOI signals from the KOI spreadsheet with the detected KOI signals from LcSignalFinder thereby producing a comparison report for study. For detected signals having periods that were multiples of the reference periods, the script normalized the period and epoch of the detected signals so that a direct comparison could be made between signals.

LcViewer was then executed to visually confirm the signals recorded by LcSignalFinder. Both the reference KOI signals and detected KOI signals were displayed together in a light curve to show any discrepancies.

If a KOI signal could not be detected, the light curve was manually detrended in an attempt to find it. The light curve was also manually detrended if the KOI signal was detected but the signal duration was too long or too short due to underfitting or overfitting the light curve initially. Once the applicable light curves were manually detrended and saved, LcSignalFinder was rerun and the process repeated as described above.

\subsubsection{Results} \label{subsec:test2results}

\begin{figure}[ht]
\includegraphics[scale=0.70]{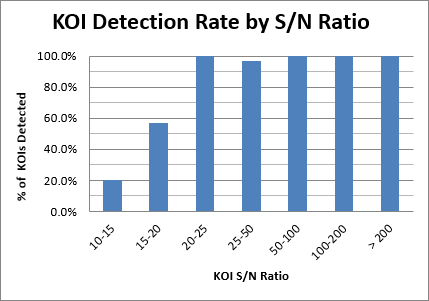}
\centering
\caption{KOI Signal Detection Rate by S/N Ratio.}
\label{fig:Test2DetectionResults}
\end{figure}

Of the 174 target KOI signals, LcSignalFinder detected 162 signals for an overall success rate of 93.1\%. The higher the S/N ratio of the reference KOI signals the higher the KOI detection rate as shown in Figure \ref{fig:Test2DetectionResults}.

Of the 157 reference KOI signals having a S/N ratio $\geq$ 20, 156 signals (99.4\%) were found. Below 20, there was a steep drop-off with only 2 out of 10 signals (20\%) found at a S/N ratio $\leq$ 15. Of the 11 signals that were not detected for a S/N ratio below 20, 10 had periods exceeding 110 days. Thus, the worst detection results were obtained for signals having a low S/N ratio combined with a very long period.

Of the 162 KOI signals found by LcSignalFinder, manual detrending of the light curve was required to detect 8 signals (4.9\%) due to underfitting or overfitting the light curve initially.

A total of 20 signals (12.3\%) had periods that were multiples of the reference KOI signals typically in association with EBs having a distinct primary and secondary eclipse signal.

\begin{figure}[ht]
\includegraphics[scale=0.67]{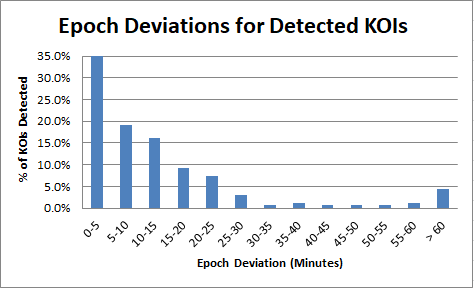}
\centering
\caption{Epoch Deviations for Detected KOI Signals.}
\label{fig:Test2EpochPrecisionResults}
\end{figure}

Figure \ref{fig:Test2EpochPrecisionResults} shows the epoch deviation in minutes between the reference KOI signals and the detected KOI signals. The vast majority of signals (71.0\%) had an epoch deviation under 15 minutes with 35.8\% of all signals having a deviation under 5 minutes.

\begin{figure}[ht]
\includegraphics[scale=0.65]{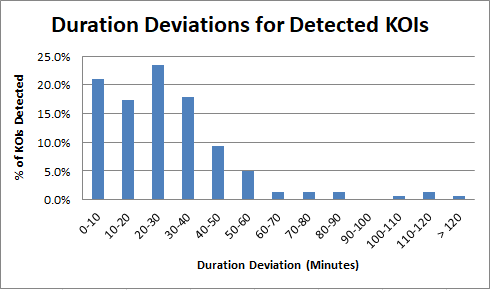}
\centering
\caption{Duration Deviations for Detected KOI Signals.}
\label{fig:Test2DurationPrecisionResults}
\end{figure}

Figure \ref{fig:Test2DurationPrecisionResults} shows the duration deviation in minutes between the reference KOI signals and the detected KOI signals. The vast majority of signals (79.6\%) had a duration deviation under 40 minutes. Of the 162 KOI signals found by LcSignalFinder, manual detrending of the light curve was performed to improve the duration for 12 signals (7.4\%) due to underfitting or overfitting the light curve initially.

\begin{figure}[ht]
\includegraphics[scale=0.67]{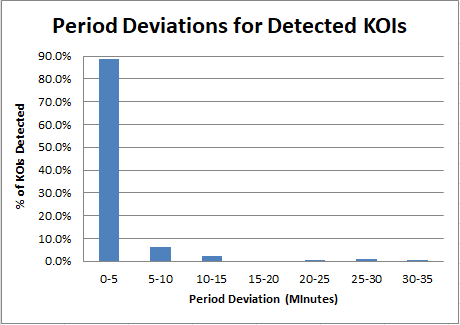}
\centering
\caption{Period Deviations for Detected KOI Signals.}
\label{fig:Test2PeriodPrecisionResults}
\end{figure}

Figure \ref{fig:Test2PeriodPrecisionResults} shows the period deviation in minutes between the reference KOI signals and the detected KOI signals. Nearly 90\% of signals had a period deviation under 5 minutes.

Of the 106 reference KOIs having a signal disposition of CONFIRMED, QuickFind assigned a signal type of PC (Planet Candidate) to 96 signals for a 90.6\% success rate at determining the correct signal type.

\subsection{Test 3 -- Detection of Single Events} \label{subsec:singleeventdetection}

\subsubsection{Objectives} \label{subsec:test3objectives}

The primary objectives of test 3 were to 1) determine the detection rate of single KOI events (standalone KOI instances) in a set of light curves over various S/N ratios, and 2) determine the deviation in epoch and duration between the reference KOI instances and the detected KOI instances.

\newpage
\subsubsection{Procedure} \label{subsec:test3procedure}

Target KOIs for the test were first selected from the KOI spreadsheet using the data filters TTV value $\leq$ 0.5 hours, S/N ratio $\geq$ 10, signal duration $\geq$ 1.5 hours, signal period $\geq$ 45 days, and signal disposition = CONFIRMED or CANDIDATE. This produced an initial list of 127 KOIs. Five KOIs were removed from the list due to various issues leaving 122 KOIs for the test.

Next, LcGenerator was executed to build Kepler long cadence PDCSAP light curves for the host stars using all available time series data. To maximize the number of data points in each light curve, the quality filter was turned off.

Using LcViewer each light curve was then manually inspected to find a time series quarter containing a single high quality KOI instance to serve as a search target for the test.

After all light curves were inspected, LcGenerator was again executed to build light curves for the test only this time using the time series quarters determined in the previous step. Each light curve produced spanned one time series quarter and contained a single KOI instance.

\begin{figure}[ht]
\includegraphics[scale=0.58]{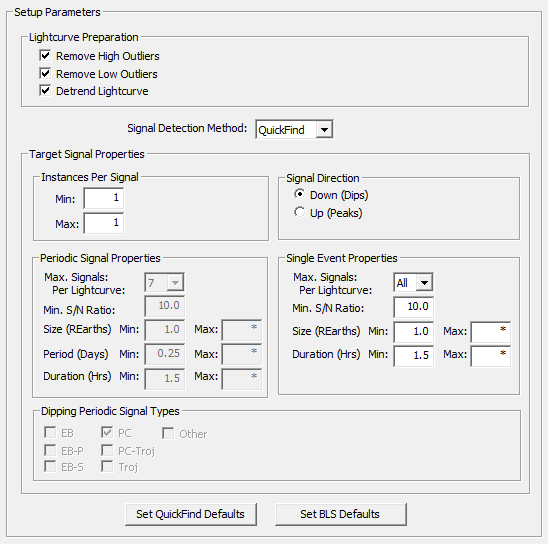}
\centering
\caption{LcSignalFinder Settings Used for Test 3.}
\label{fig:LcSignalFinderTest3Settings}
\end{figure}

LcSignalFinder was then executed on the light curves produced by LcGenerator using the settings shown in Figure \ref{fig:LcSignalFinderTest3Settings}. Prior to starting the job, the signal library for KOIs was deselected thereby enabling LcSignalFinder to detect KOI signals.

Upon completion, LcReporter was executed to generate an Excel report listing the signals detected by LcSignalFinder. A custom script was then run which combined the reference KOI signals from the KOI spreadsheet with the detected KOI signals from LcSignalFinder thereby producing a comparison report for study. If the reference epoch for a KOI signal was for different instance relative to the detected instance, the reference epoch was normalized so that a direct comparison could be made between instances.

LcViewer was then executed to visually confirm the signals recorded by LcSignalFinder. Both the reference KOI instance and detected KOI instance were displayed together in a light curve to show any discrepancies.

If the KOI instance could not be detected, the light curve was manually detrended in an attempt to find it. The light curve was also manually detrended if the KOI instance was detected but the signal duration was too long or too short due to underfitting or overfitting the light curve initially. Once the applicable light curves were manually detrended and saved, LcSignalFinder was rerun and the process repeated as described above.

\subsubsection{Results} \label{subsec:test3results}

\begin{figure}[ht]
\includegraphics[scale=0.70]{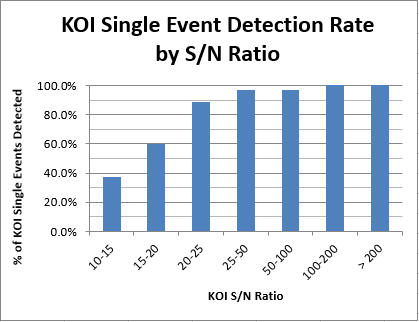}
\centering
\caption{KOI Single Event Detection Rate by S/N Ratio.}
\label{fig:Test3DetectionResults}
\end{figure}

Of the 122 target KOI instances, LcSignalFinder detected 112 instances for an overall success rate of 91.8\%. Like periodic signals in the previous test, the higher the S/N ratio of the reference KOI signals the higher the detection rate of single instances as shown in Figure \ref{fig:Test3DetectionResults}.

Of the 109 reference KOI instances having a S/N ratio $\geq$ 20, 106 instances (97.2\%) were found. Below 20, there was a steep drop-off with only 3 out of 8 signals (37.5\%) found at a S/N ratio $\leq$ 15.

Of the 112 KOI instances found by LcSignalFinder, manual detrending of the light curve was required to detect 13 instances (11.6\%) due to overfitting the light curve initially.

\begin{figure}[ht]
\includegraphics[scale=0.70]{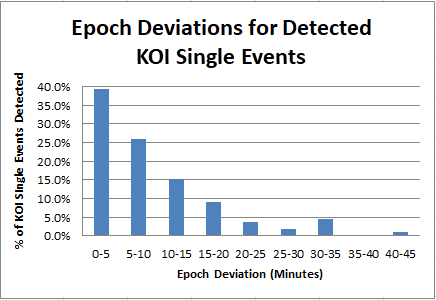}
\centering
\caption{Epoch Deviations for Detected KOI Single Events.}
\label{fig:Test3EpochPrecisionResults}
\end{figure}

Figure \ref{fig:Test3EpochPrecisionResults} shows the epoch deviation in minutes between the reference KOI instances and the detected KOI instances. The vast majority of instances (80.4\%) had an epoch deviation under 15 minutes with 39.3\% of all instances having a deviation under 5 minutes.

\begin{figure}[ht]
\includegraphics[scale=0.70]{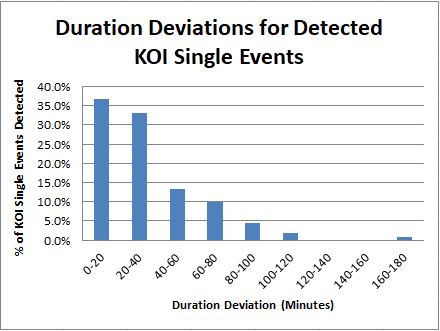}
\centering
\caption{Duration Deviations for Detected KOI Single Events.}
\label{fig:Test3DurationPrecisionResults}
\end{figure}

Figure \ref{fig:Test3DurationPrecisionResults} shows the duration deviation in minutes between the reference KOI instances and the detected KOI instances. The vast majority of instances (69.6\%) had a duration deviation under 40 minutes with 36.6\% of all instances having a deviation under 20 minutes. Of the 112 KOI instances found by LcSignalFinder, manual detrending of the light curve was performed to improve the duration for 16 instances (14.3\%) due to overfitting the light curve initially.

\subsection{Test 4 -- Detection of Periodic Signals With TTVs} \label{subsec:ttvsignaldetection}

\subsubsection{Objectives} \label{subsec:test4objectives}

The primary objectives of test 4 were to 1) determine the detection rate of KOI signals with strong TTVs in a set of light curves, 2) determine the deviation in epochs between the reference KOI instances and the detected KOI instances after TTV offsets were applied, and 3) determine the deviation in signal duration and period between the reference KOI signals and the detected KOI signals.

\subsubsection{Procedure} \label{subsec:test4procedure}

An initial set of target KOIs for the test were first selected from the KOI spreadsheet using the data filters Kepler quarters observed $\geq$ 11, TTV value $\geq$ 0.65 hours, S/N ratio $\geq$ 25, signal duration $\geq$ 2.5 hours, signal period $\leq$ 165 days, and signal disposition = CONFIRMED or CANDIDATE.

LcGenerator was then executed to build Kepler long cadence PDCSAP light curves for the host stars using all available time series data. To maximize the number of data points in each light curve, the quality filter was turned off.

Next, LcViewer was executed to manually inspect the light curves for viable KOI signals to be used in the test. TTV offsets from the Holczer TTV catalog \citep{Holczer_2016} were applied to the KOI instances to obtain the observed instance positions. A KOI signal was deemed viable for the test if the instances were easily visible in the light curve, had at least three instances with large TTVs, and produced a TTV diagram with a relatively smooth TTV curve similar to that shown in Figure \ref{fig:TTVDiagram}. From the initial set of target KOIs, 23 viable KOI signals were selected containing 888 individual instances.

\begin{figure}[ht]
\includegraphics[scale=0.58]{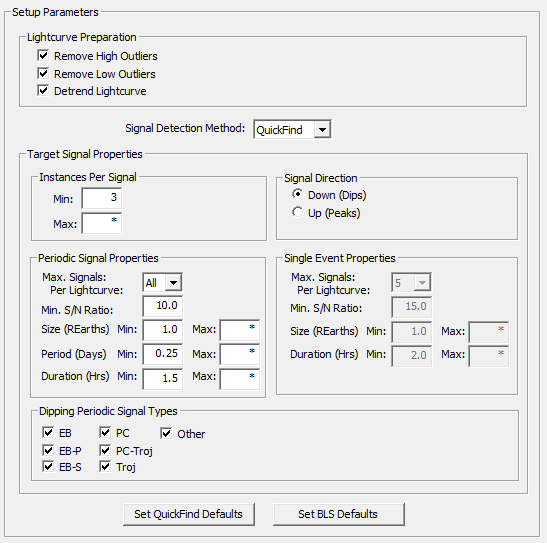}
\centering
\caption{LcSignalFinder Settings Used for Test 4.}
\label{fig:LcSignalFinderTest4Settings}
\end{figure}

LcSignalFinder was then executed on the light curves for the selected KOIs using the settings shown in Figure \ref{fig:LcSignalFinderTest4Settings}. Prior to starting the job, the signal library for KOIs was deselected thereby enabling LcSignalFinder to detect KOI signals.

Upon completion, LcReporter was executed to generate an Excel report listing the signals and TTVs detected by LcSignalFinder. A custom script was then run which combined the reference KOI signals from the KOI spreadsheet with the detected KOI signals from LcSignalFinder thereby producing a comparison report for study. For both sets of signals, TTV offsets were applied to instances so that observed positions could be compared.

Next, LcViewer was executed to visually confirm the signals and TTVs recorded by LcSignalFinder. Both the reference KOI signals and detected KOI signals were displayed together in a light curve to show any discrepancies. TTV offsets were applied to instances to show the observed positions. TTV diagrams were also produced to identify outliers indicating inaccurate placement of instances.

If a KOI signal could not be detected, the light curve was manually detrended in an attempt to find it. The light curve was also manually detrended if the KOI signal was detected but the signal duration was too long or too short due to underfitting or overfitting the light curve initially. Once the applicable light curves were manually detrended and saved, LcSignalFinder was rerun and the process repeated as described above.

\subsubsection{Results} \label{subsec:test4results}

LcSignalFinder successfully detected all 23 target KOI signals and 885 out of 888 target KOI instances (99.7\%). Manual detrending of the light curve was required to detect 3 signals (13\%) due to underfitting or overfitting the light curve initially.

\begin{figure}[ht]
\includegraphics[scale=0.66]{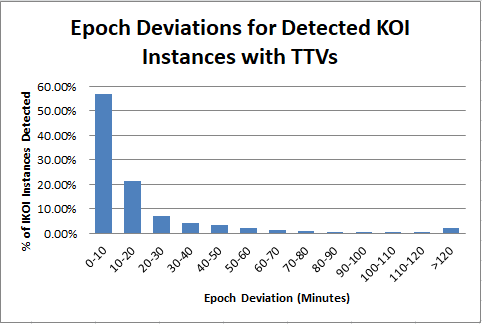}
\centering
\caption{Epoch Deviations for Detected KOI Instances with TTVs.}
\label{fig:Test4EpochPrecisionResults}
\end{figure}

Figure \ref{fig:Test4EpochPrecisionResults} shows the epoch deviation in minutes between the reference KOI instances and the detected KOI instances after TTV offsets were applied. Of the 885 KOI instances found, 502 (56.7\%) had a epoch deviation under 10 minutes and 691 instances (78.1\%) had an epoch deviation under 20 minutes.

\begin{figure}[ht]
\includegraphics[scale=0.62]{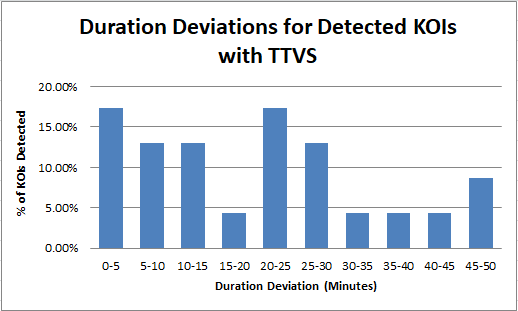}
\centering
\caption{Duration Deviations for Detected KOI Signals with TTVs.}
\label{fig:Test4DurationPrecisionResults}
\end{figure}

Figure \ref{fig:Test4DurationPrecisionResults} shows the duration deviation in minutes between the reference KOI signals and the detected KOI signals. The vast majority of signals (78.3\%) had a duration deviation under 30 minutes.

\begin{figure}[ht]
\includegraphics[scale=0.62]{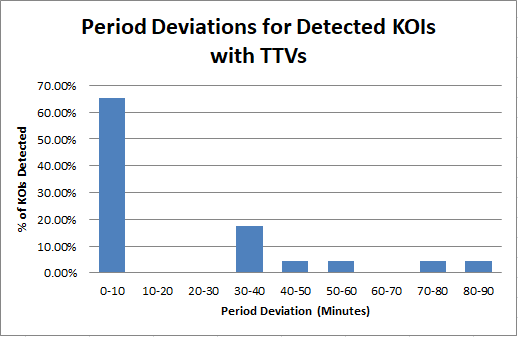}
\centering
\caption{Period Deviations for Detected KOI Signals with TTVs.}
\label{fig:Test4PeriodPrecisionResults}
\end{figure}

Figure \ref{fig:Test4PeriodPrecisionResults} shows the period deviation in minutes between the reference KOI signals and the detected KOI signals. The vast majority of signals (65.2\%) had a period deviation under 10 minutes.

Of the 16 reference KOIs having a signal disposition of CONFIRMED, QuickFind assigned a signal type of PC (Planet Candidate) to all of them for a 100\% success rate at determining the correct signal type.


\section{Summary} \label{sec:summary}

This paper described the new QuickFind method in LcTools for detecting signals and associated TTVs in light curves. Sections \ref{sec:finder} thru \ref{sec:reporter} covered the changes made to LcSignalFinder, LcViewer, and LcReporter to support QuickFind and related operations. Section \ref{sec:finder} also provided a detailed description of the QuickFind data processing algorithm used in LcSignalFinder.

Section \ref{sec:performance} covered the QuickFind performance tests that were conducted in LcSignalFinder to assess processing speed, signal detection and precision, and TTV detection and precision.

Test 1 measured processing speed for 425 light curves across various timespans. QuickFind proved to be extremely fast at all timespans. The average execution time per light curve ranged from 0.16 seconds at 33 days to 4.72 seconds at 1,459 days. Relative to BLS, QuickFind was much faster -- 2.75 times faster at 33 days and 76 times faster at 1,459 days.

Test 2 determined the ability of QuickFind to detect 174 target KOI signals in a set of 167 light curves over various S/N ratios. QuickFind found 99.4\% of all KOI signals having a S/N ratio above 20 but only 35.3\% of all signals having a S/N ratio below 20. Deviations between the reference KOI signals and the detected KOI signals were generally below 15 minutes for epoch, below 40 minutes for duration, and below 5 minutes for period.

Test 3 determined the ability of QuickFind to detect 122 target KOI single events (standalone KOI instances) in a set of 122 light curves over various S/N ratios. QuickFind found 97.2\% of all single events having a S/N ratio above 20 but only 46.2\% of all single events having a S/N ratio below 20. Deviations between the reference KOI instances and the detected KOI instances were generally below 15 minutes for epoch and below 40 minutes for duration.

Test 4 determined the ability of QuickFind to detect 23 target KOI signals having strong TTVs in a set of 19 light curves. QuickFind found 100\% of the KOI signals and 99.7\% of the individual KOI instances after TTV offsets were applied. Deviations between the reference KOI signals and the detected KOI signals were generally below 20 minutes for epoch, below 30 minutes for duration, and below 10 minutes for period.

The single greatest factor affecting signal detection and precision was automatic detrending of light curves on the initial processing pass. Due to the effects of underfitting and overfitting, some target signals were missed or had signal durations that were too long or too short. Manual detrending was required on the second processing pass to find 8.1\% of KOI signals and to improve the duration for 9.4\% of signals.

Collectively, the four tests demonstrate the viability of QuickFind in detecting signals and TTVs in light curves. Together with BLS, there are now two distinct methods in LcTools for finding signals, each with advantages and disadvantages.

BLS is strong at finding signals of all sizes including those with S/N ratios below 20 where QuickFind has more difficulty. On the flip side, BLS incurs a heavy cost in processing time especially for light curves over 300 days.

By contrast, QuickFind is adept at finding medium to large sized signals (generally those with S/N ratios above 15) extremely fast regardless of the timespan. Unlike BLS, it can also detect signals with TTVs.

Ultimately, the LcTools user must select the most appropriate method based on the number of light curves in the job, the timespan of the light curves, and the smallest signal size to be found. Many individuals in the LcTools community currently use LcSignalFinder to process extremely large sets of TESS light curves involving hundreds of thousands or even millions of files, some spanning multiple sectors with long timespans. Due to very fast signal processing, QuickFind is generally considered the method of choice for most people.

The LcTools software system is free and can be obtained from the lead author by request at \\ \href{mailto:aschmitt@comcast.net}{aschmitt@comcast.net}. For additional information on the product, see the LcTools Product Description\footnote{\url{https://sites.google.com/a/lctools.net/lctools/lctools-product-description}}.


\newpage
\appendix

\section{Run-Time Requirements} \label{sec:rtreqs}

The following hardware and software is required to run LcTools:
\begin{itemize}
\item Windows OS (7/8/10).
\item 2.7 GHz CPU.
\item 5 GB memory.
\item 50 GB free disk space for light curve files and associated data files on any drive. 20 MB for the LcTools software system on drive C.
\item 1024 x 768 screen resolution.
\item 2-Button mouse or equivalent.
\item High-speed Internet connection.
\item Microsoft Word.
\item Microsoft Excel (if generating Excel reports).
\item Google Drive (if using public signal libraries or public TTV libraries).
\end{itemize}

\acknowledgments

\noindent A sincere thanks is extended to the following organizations and individuals for their contributions to this paper:

To MAST for providing access to light curves, stellar properties, and signals for the TESS, K2, and Kepler missions and associated High Level Science Products.

To the NASA Exoplanet Archive (NEA) for providing access to stellar properties and signals for the K2 and Kepler missions.

To ExoFOP-TESS\footnote{\url{https://exofop.ipac.caltech.edu/tess/}} for providing access to TOIs and CTOIs for the TESS mission.

To Tomer Holczer for use of his TTV database for aligning KOI signals.

To Scott Fleming for providing technical assistance in accessing data at MAST.

To Robert Gagliano, Tom Jacobs, Martti Kristiansen, Daryll LaCourse, and Hans Schwengeler for their assistance in testing LcTools and contributing ideas for new features.

To Joel Hartman for reviewing this paper and providing helpful comments.



\bibliographystyle{aasjournal}


\begin{figure}[ht]
\includegraphics[scale=0.75]{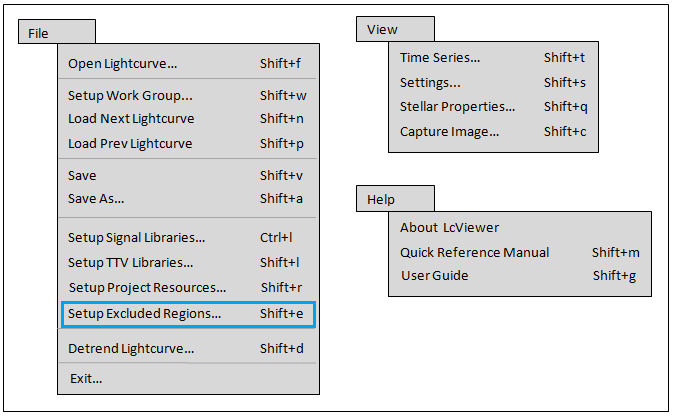}
\centering
\caption{Changes to Menu Bar Commands in LcViewer (Blue = Added).}
\label{fig:MenuBar}
\end{figure}

\begin{figure}[ht]
\includegraphics[scale=0.85]{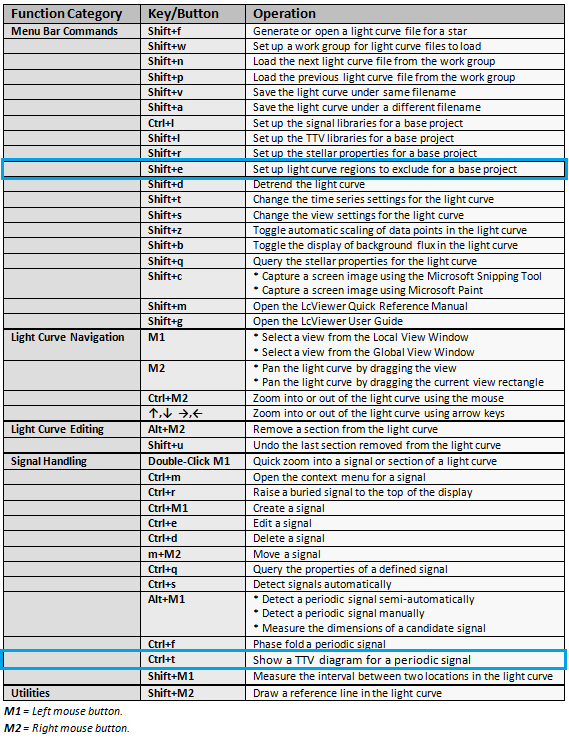}
\centering
\caption{Changes to Hot Keys and Hot buttons in LcViewer (Blue = Added).}
\label{fig:Hotkeys}\end{figure}


\end{document}